\documentclass[12pt,preprint]{aastex}

\shorttitle{72 DA White Dwarfs Identified in LAMOST Pilot Survey }

\shortauthors{Zhao et al.}

\begin{document}


\title{\bf 72 DA White Dwarfs Identified in LAMOST Pilot Survey}


\author{J. K. Zhao\altaffilmark{1}, A. L. Luo\altaffilmark{1}, T.D. Oswalt\altaffilmark{2}, G. Zhao\altaffilmark{1}}


\altaffiltext{1}{Key Laboratory of Optical Astronomy, National Astronomical Observatories, Chinese Academy of Sciences, Beijing, 100012, China; zjk@bao.ac.cn, gzhao@bao.ac.cn, lal@bao.ac.cn}

\altaffiltext{2}{Physics and Space Science Department, Florida Institute of Technology, Melbourne, USA, 32901, toswalt@fit.edu}


\begin{abstract}
We present a spectroscopically identified  catalogue of 72 DA white dwarfs from the LAMOST pilot survey. 35 are found to be new identifications after cross-correlation with the Eisenstein et al. and Villanova catalogues. The effective temperature and gravity of these white dwarfs are estimated by Balmer lines fitting. Most of them are hot white dwarfs. The cooling times and masses of these white dwarfs are estimated by interpolation in theoretical evolution tracks. The peak of mass distribution is found to be $\sim$ 0.6 $M_\odot$ which is consistent with prior work in the literature. The distances of these white dwarfs are estimated using the method of Synthetic Spectral Distances. All of these WDs are found to be in the Galactic disk from our analysis of space motions. Our sample supports the expectation white dwarfs with high mass are concentrated near the plane of Galactic disk.   
\end{abstract}
\keywords{white dwarfs: Stars}



\section{Introduction}
White dwarfs (WDs) are the final stage for the the evolution of majority
of low and medium mass stars with initial masses $<$ $8 M_\odot$. Since there are no fusions reaction, the evolution of WDs is primarily determined by a well understood
cooling process (Fontaine et al. 2001; Salaris et al.
2000). Thus, they can be used for cosmochronology, an independent age-dating method. Also, the luminosity function of WDs provides firm constraints on the local star formation rate
and history of the Galactic disk (Krzesinski et al. 2009).

 McCook $\&$ Sion (1999) present a catalog of 2249 WDs which have been identified spectroscopically. In addition, the Sloan Digital Sky Survey (SDSS; York et al. 2000) has greatly expanded the number of spectroscopically confirmed WD stars (Harris et al. 2003; Kleinman et al. 2004; Eisenstein et al. 2006; Kleinman et al. 2013). The latter presented a catalog of 20,407 spectroscopically confirmed white dwarfs from the SDSS Data Release 4 (DR4), roughly doubling the number of spectroscopically
confirmed white dwarfs.

Large sky Area Multi-Object Fiber Spectroscopic Telescope (LAMOST, so called the Guoshoujing Telescope) is a National Major Scientific Project undertaken by the Chinese Academy of Science (Wang et al. 1996; Cui et al. 2012). LAMOST has recently completed the pilot survey from October 2011 to May 2012, which obtained several hundred thousand spectra (Luo et al. 2012). From September of 2012, LAMOST has undertaken the general survey and will observe about 1 million stars per year. LAMOST has the capability to observe large, deep and dense regions in the Milky Way Galaxy, which will enable a number of research topics to explore the evolution and the structure of the Milky Way. Therefore, it will definitely yield a large sample of WDs.

WDs whose primary spectral classification is DA have hydrogen-dominated atmospheres. They make up the majority (approximately 75\%) of all observed WDs (Fontaine $\&$ Wesemael 2001). Such WDs are easy to identify using optical spectra.  Here we present a catalog of DA WDs from the LAMOST pilot survey (Luo et al. 2012). {\bf We do not expect the completeness of this sample.} In section 2 we describe the spectra obtained. Section 3  discusses how the $T\rm_{eff}$, log $g$, mass and distance of the WDs were estimated. The kinematics of these WDs are illustrated in section 4. A summary of our pilot study results is given in section 5.



\section{LAMOST Pilot Data and Observations}
The LAMOST spectra have a resolving power of  R $\sim$ 2000 spanning 3700$\rm \AA \sim $ 9000$\rm\AA$. Two arms of each spectrograph cover this wavelength range with overlap of 200 $\rm \AA$. The blue spectral coverage is  3700$\rm \AA \sim$ 5900$\rm\AA$ while that in the red is 5700$\rm \AA \sim$ 9000 $\rm\AA$. The raw data were reduced with LAMOST 2D and 1D pipelines (Luo et al. 2004). These pipelines include bias subtraction, cosmic-ray removal, spectral trace and extraction, flat-fielding, wavelength calibration, sky subtraction, and combination. The throughput in red is higher than the blue band.

 The pilot survey obtained spectra of stars in the Milky Way, which included fainter objects on dark nights (Yang et al. 2012; Carlin et al. 2012), brighter objects on bright nights (Zhang et al. 2012), objects in the disk of the Galaxy with low latitude (Chen et al. 2012) and objects in the region of the Galactic Anti-Center. It also targets extragalactic objects located in two regions, i.e., the South Galactic Cap and the North Galactic Cap.

 We found twenty WD spectra in both SDSS and LAMOST pilot survey catalogs. Fig. 1 shows a portion of a typical spectrum. The top panel compares the SDSS DR7 and LAMOST spectra for the object J100316.35-002336.95. The solid line is the SDSS spectrum. The dotted line is the LAMOST spectrum. The bottom panel shows the residual between two spectra. The mean difference between two spectra is less than 10 \%.

 The initial WD candidates we selected are from two sources. One is the LAMOST pipeline (Luo et al. 2012) which yielded about 2000 candidates using the "PCAZ" method. For stars with SDSS photometry, we used the formulas 1-4 of  Elsenstein et al. (2006) to idendify candidates. Next, each of these spectra was inspected by eye. Stars with signal-to-noise ratio (S/N) smaller than 10 were excluded. Finally, if the Balmer line profiles of the star were a little too narrow (log $g$ $<$ 7.0), the spectrum was rejected even if selected by the pipeline. After these filters, 72 DA WDs were left. Table 1 presents the physical data for these WDs. Column 1 is an ID number. Columns 2-5 list the name, RA and DEC. The estimated $T\rm_{eff}$, log $g$, mass and the cooling time are given in columns 5-8. Columns 9-13 list the apparent magnitudes of each WDs. Column 14 indicates the source of the magnitudes. The last two columns are estimates of the color excess (B-V) and distance. The E(B-V) is estimated from Schlegel et al. (1998).

\section{Parameter Determination}
\subsection{$T\rm_{eff}$ and log $g$}
For our DA WD candidates, the $T$$\rm_{eff}$ and log $g$ were derived via simultaneous
fitting of the H$\beta$ to H8 Balmer line profiles using the procedure outlined by Bergeron
et al. (1992). The line profiles in both observed spectra and model spectra were
first normalized using two points at the continuum level on either
side of each absorption line. Thus, the fit should not be
affected by the flux calibration. Model atmospheres used for this fitting were derived
from model grids provided by Koester (2010). Details of the input physics and
methods can be found in that reference. Fitting of the line profiles
was carried out using the IDL package MPFIT (Markwardt 2008), which is based on
$\chi^{2}$ minimization using Levenberg-Marquardt method. This package
can be downloaded from the project website\footnote[1]{http://purl.com/net/mpfit}.
Errors in the $T$$\rm_{eff}$ and log $g$ were calculated by
stepping the parameter in question away from their optimum values
and redetermining minimum $\chi^{2}$ until the difference between this
and the true minimum $\chi^{2}$ corresponded to 1$\sigma$ for a given number of
free model parameters.

Figs. 2-3 show examples of $T$$\rm_{eff}$ and log $g$ determinations for J150156.26+302300.13. Fig.2 is the contour plot of the $\chi^{2}$ residual and the rough $T$$\rm_{eff}$ and log $g$ implied by these error eclipses. Fig. 3 shows the actual fits of the observed Balmer lines for J150156.26+302300.13. The black solid lines are the observed profiles of Balmer lines from H$\beta$ to H8. The red dashed lines are the model spectra. The derived $T$$\rm_{eff}$, log $g$ and uncertainties for all the WDs are shown in columns 5-6 of Table 1. Estimated $T$$\rm_{eff}$ and log $g$ values for 14 DAs were also available in the literature, allowing the comparisons shown in Fig. 4. The solid line represents the unit slope relation. Plus (+) symbols represent the WDs with high S/N spectra while squares represent WDs with low S/N spectra. The three spectra of lowest S/N are outliers in the log $g$ comparison plot-suggesting the importance of S/N in determining this parameter. For most of other WDs, the mean differences between our and the literature $T$$\rm_{eff}$ values are less than 1000 K and the log $g$ difference is less than 0.2 dex. Within this scatter, our results are consistent with those in the literatures. One of our candidates, J104311.45+490224.35 has also been identified as DA WD by McCook $\&$ Sion (1999). However, we were unable to determine its $T$$\rm_{eff}$ and log $g$ because H$\beta$ was not included in the spectrum we obtained.

\subsection{Mass and Cooling Time}
From the $T$$\rm{_{eff}}$ and log $g$ of each WD, its mass (M$\rm_{WD}$) and cooling time (t$\rm_{cool}$) were estimated from Bergeron's cooling sequences\footnote[2]{The cooling sequences can be downloaded from the website: http://www.astro.umontreal.ca/$\sim$bergeron/CoolingModels/.}. For the model atmospheres above $T$$\rm{_{eff}}$ = 30,000 K we used the carbon-core cooling models of Wood (1995), with thick hydrogen layers of q$\rm_{H}$ = M$\rm_{H}$/M$_{*}$ = 10$^{-4}$. For $T$$\rm{_{eff}}$   below 30,000 K we used cooling models similar to those described in Fontaine, Brassard $\&$ Bergeron (2001) but with carbon-oxygen cores and q$\rm_{H}$ = 10$^{-4}$ (see Bergeron, Leggett $\&$ Ruiz 2001).

Fig. 5 is the mass distribution of our sample resulting from the above procedure. Masses are found to range from 0.4 $M_\odot$ to 1.2 $M_\odot$. The curve is a Gaussian fit with a peak at about 0.61 $M_\odot$, which is consistent with the mean mass 0.613 $M_\odot$ from Tremblay et al. (2011) derived from SDSS DA WDs sample.

\subsection{Distance}
The determination of distances for WDs is  very difficult because of their low luminosity. Currently only about 300 WDs have trigonometric parallaxes. In the absence of parallaxes, color-magnitude relations and empirical photometric methods are often used.  Holberg et al. (2008) provided improved distance estimates for DA WDs
using multi-band synthetic photometry tied to spectroscopic temperatures and gravities. This method was called Synthetic Spectral Distances (SSD). 
The unique aspect of SSD is the systematic use of calibrated multi-channel synthetic absolute magnitudes, interpolated within the grid by the $T\rm_{eff}$ and log $g$.

\begin{eqnarray}
m_{i}&=  &\sum_{i=(u,g,r,i,z,V)}M_{i}(log g,T_{eff})+a_{i}A_{g}+5logd-5
\end{eqnarray}

In this paper, the distances of WDs in our sample were estimated using Equation 1. Here, $m_{i}$ are the photometric magnitudes of the WDs. Most of our has \textit{u}, \textit{g}, \textit{r}, \textit{i}, \textit{z} magnitudes. Almost all have at least \textit{g}, \textit{r} and \textit{i} magnitudes. A few WDs still only have \textit{V} magnitude. $M_{i}$ is the model absolute magnitudes calculated by interpolations in the atmospheric models provided by Bergeron.  $A_{g}a_{i}$ is the reddening and d is the distance in parsecs. In general for each magnitude a corresponding distance can be calculated. The final distance is estimated by using weighted average. The weights adopted are the errors in the magnitude. Here, we only calculated the distances for WDs having \textit{u}, \textit{g}, \textit{r}, \textit{i}, \textit{z} or \textit{V} magnitude data. Distances for two WDs in Table 1 could not be estimated.
\section{Kinematics}
Oppenheimer et al. (2001) suggested that halo WDs could provide a significant contribution to the Galactic dark matters component, which prompted much interest in WD kinematics. In a related study, Silvestri et al. (2002) observed 116 common proper-motion binaries
consisting of a WD plus M dwarf component. They determined full space motions of their WDs from the companion M dwarfs. Most of their WDs were found to be members of the disk; only one potential halo WD was identified. Even the much larger samples of WDs such as the
Pauli et al. (2003, 2006) SN Ia Progenitor Survey (SPY) have found relatively few genuine halo and thick disk candidates. In their magnitude-limited sample of 398 WDs, they examined both the
UVW space motions and the Galactic orbits of their stars. They
found only 2\% of their sample kinematically belonged to
the halo and 7\% to the thick disk.
Sion et al. (2009) presented the kinematical properties of the WDs within 20 pc of the Sun. In their nearby sample, they found no convincing evidence of halo members among 129 WDs, nor was there convincing evidence of genuine thick disk subcomponent members within
20 parsecs. The entire 20 pc sample likely belongs to the thin disk.

 The proper motions of our sample were derived by the cross-correlating with PPMXL catalog (Roeser et al. 2010). Silvestri et al. (2002); Pauli et al. (2003, 2006), Sion et al. (2009) found relatively little kinematical difference among the samples whether they used radial velocity (RV) to compute full space motions or used the simple zero RV for simple WDs. We have assumed zero RVs in the analysis of our sample.  U is measured
positive in the direction of the Galactic anti-center, V is measured positive in the direction of the Galactic rotation, and W is
measured positive in the direction of the north Galactic pole. The U, V and W velocities were corrected for the peculiar solar motion
(U, V, W) = (-9, +12, +7) km s$^{-1}$ (Wielen 1982). The space motions of 59 WDs with sufficient kinematical information (photometric or trigonometric parallax, proper motion) in our sample were calculated.

The top panel of Fig. 6 shows contours, centered at
(U, V) = (0, -220) km s$^{-1}$, that represent 1$\sigma$ and 2$\sigma$ velocity ellipsoids
for stars in the Galactic stellar halo as defined by Chiba $\&$
Beers (2000).  Only one of our candidate WDs lies outside the 2$\sigma$ velocity contour centered on (U, V) = (0, -35) km s$^{-1}$ defined for disk stars (Chiba $\&$ Beers 2000). The bottom of Fig. 6 shows a Toomre diagram for our stars. Venn et al. (2004) suggested stars with V$_{total}$ $>$ 180 km s$^{-1}$ are possible halo members. None of our stars  meet this criterion.
We conclude that our sample consistes entirely of disk stars.

Wegg $\&$ Phinney (2012) concluded that kinematical dispersion decreases with increasing WD mass among young WDs whose cooling time is smaller than 3$\times$10$^{8}$ years.  Progenitors of high mass WDs have shorter lifetimes, hence they should be closer to the Galactic plane and have small kinematical dispersion in accord with the disk `heating' theory. Since most WDs in our sample are relative young, we investigated the relation between mass and W, as well as mass and vertical distance of Galactic plane $|$Z$|$ (see Fig. 7). In the top panel of Fig. 7, WDs with mass larger than 0.8 $M_{\odot}$ are seen to have smaller W. Also, the vertical distances from the Galactic plane of WDs with larger mass are relative small.  Although there is no strict relation such as seen in Wegg $\&$ Phinney (2012), our sample support the general expectation that high mass WDs tend to have lower W and $|$Z$|$.

\section{Conclusions}
From the LAMOST pilot survey data, 72 DA WDs were detected with S/N $>$ 10. $T$$\rm_{eff}$, log $g$, cooling time, mass and distance of these WDs were determined from their spectra. The $T\rm_{eff}$ of most WDs range from 12000 K to 35000 K and the cooling times of all the WDs are younger than 300 Myr.  All these WDs were found to be members of Galactic disk. WDs with higher mass tend to have smaller vertical distance from the Galactic disk, which partly supports the conclusions of Wegg et al. (2012).

The DA WD catalogue of the LAMOST pilot survey provides a first glimpse of how useful the survey will be to search for nearby WDs.
 The upcoming formal LAMOST survey will
enlarge the sample of WDs rapidly, perhaps providing the largest sample of WDs available. This large sample will open the door to much more detailed investigation of the physical $\&$ kinematic properties of WDs in the solar neighborhood as well as the local structure and evolution of the Galaxy.   %




\acknowledgments
Many thanks to D. Koester for providing his WD models. Balmer/Lyman lines in the models were calculated with the modified
Stark broadening profiles of Tremblay $\&$ Bergeron (2009), kindly made available by the authors. This study  is supported by the
National Natural Science Foundation of China under grant No. 11233004, 11078019 and 10973021. T.D.O. acknowledges support from NSF grant AST-0807919 to Florida Institute of Technology.
Guoshoujing Telescope (the Large Sky Area Multi-Object Fiber Spectroscopic Telescope LAMOST) is a National Major Scientific Project built by the Chinese Academy of Sciences. Funding for the project has been provided by the National Development and Reform Commission. LAMOST is operated and managed by the National Astronomical Observatories, Chinese Academy of Sciences.

\clearpage




\begin{figure}
\epsscale{1.0}
\plotone{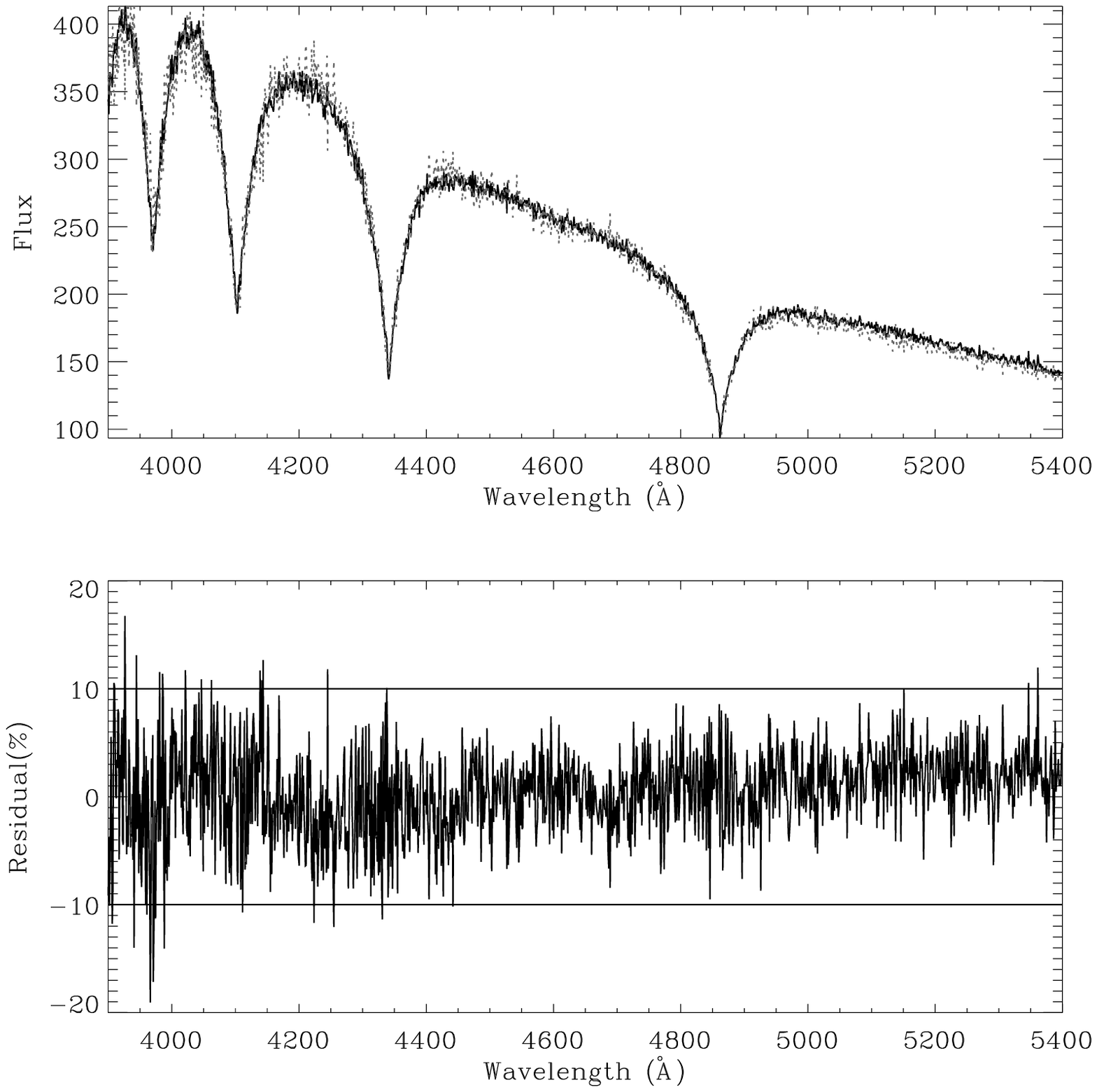}
\caption{A comparison of typical SDSS and LAMOST pilot survey WD spectra for star J100316.35-002336.95. In the top panel, the solid line is the LAMOST spectrum while dotted line is that from SDSS. The bottom panel presents the residule between the two spectra. }
\end{figure}

\begin{figure}
\epsscale{1.0}
\plotone{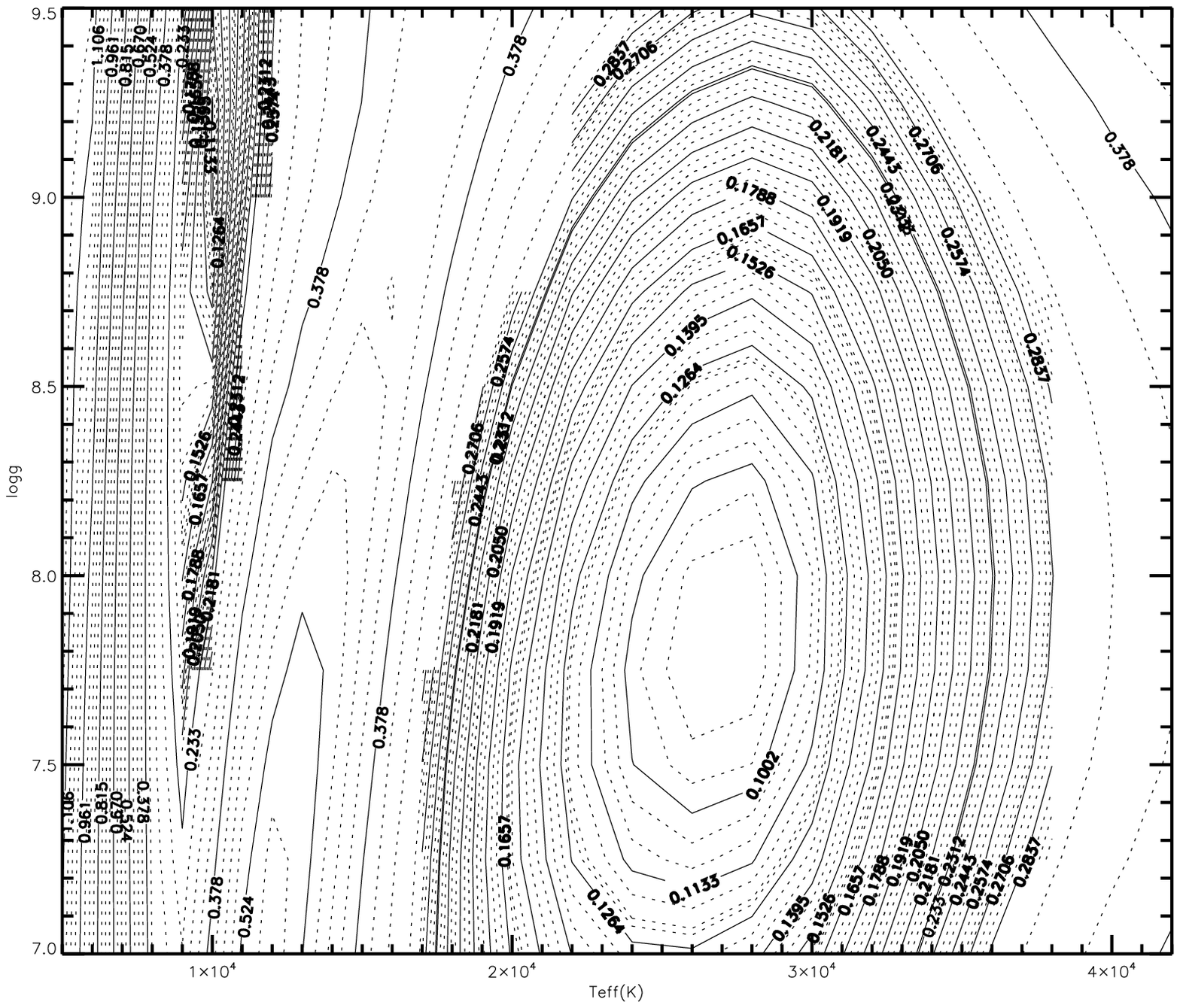}
\caption{The $\chi^{2}$ contour plot of $T\rm_{eff}$ and log $g$ determination.  }
\end{figure}

\begin{figure}
\epsscale{1.0}
\plotone{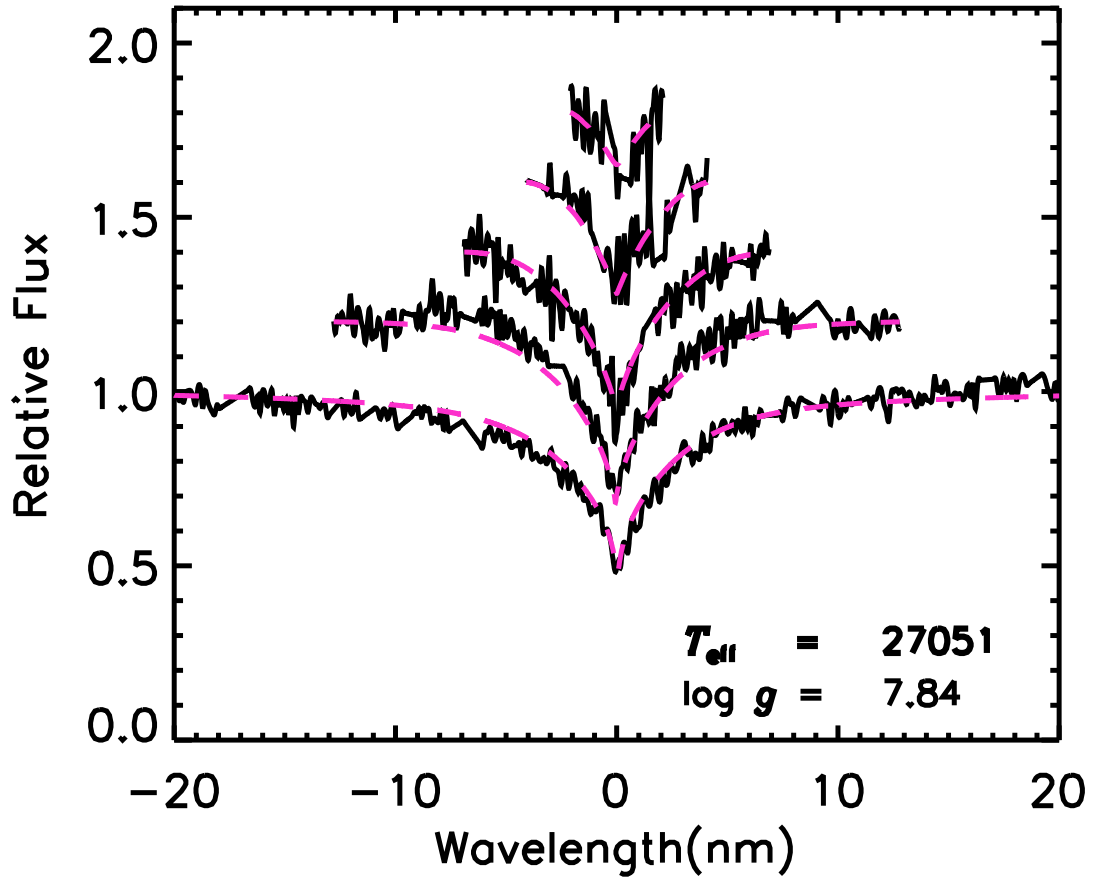}
\caption{Fits of the observed Balmer lines for J150156.26+302300.13. Lines range from H$\beta$ (bottom) to H8 (top). The solid black line is the observed spectra while the dashed line is the model spectra. }
\end{figure}

\begin{figure}
\epsscale{1.0}
\plotone{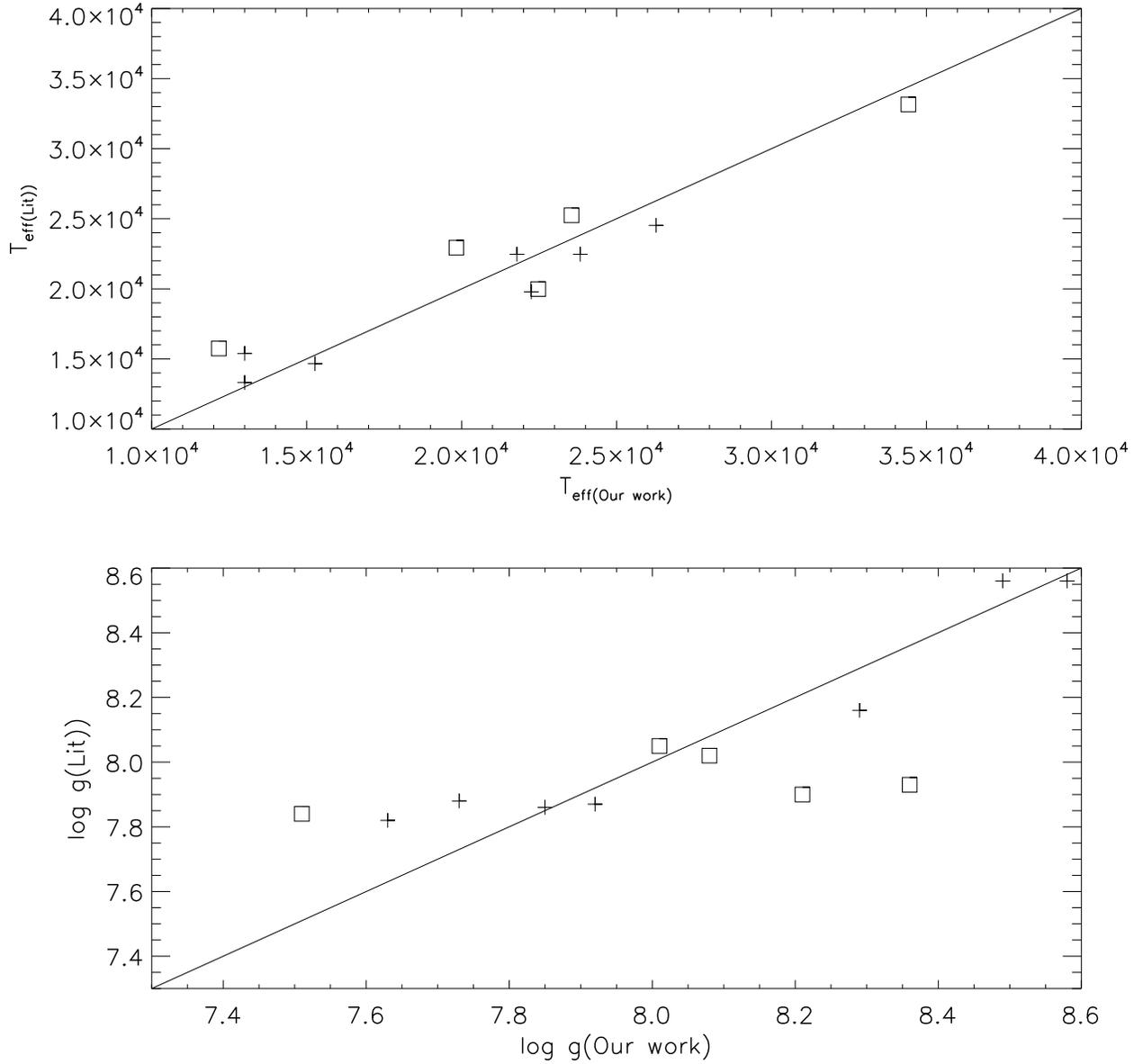}
\caption{Comparison of estimated $T\rm_{eff}$ and log $g$ values determined in this study to those from the literature. Pluses (+) represent WDs with high S/N ($>$20), while squares represent WDs with low S/N ($<$20) spectra. The
solid line is the unit slope relation. }
\end{figure}

\begin{figure}
\epsscale{1.0}
\plotone{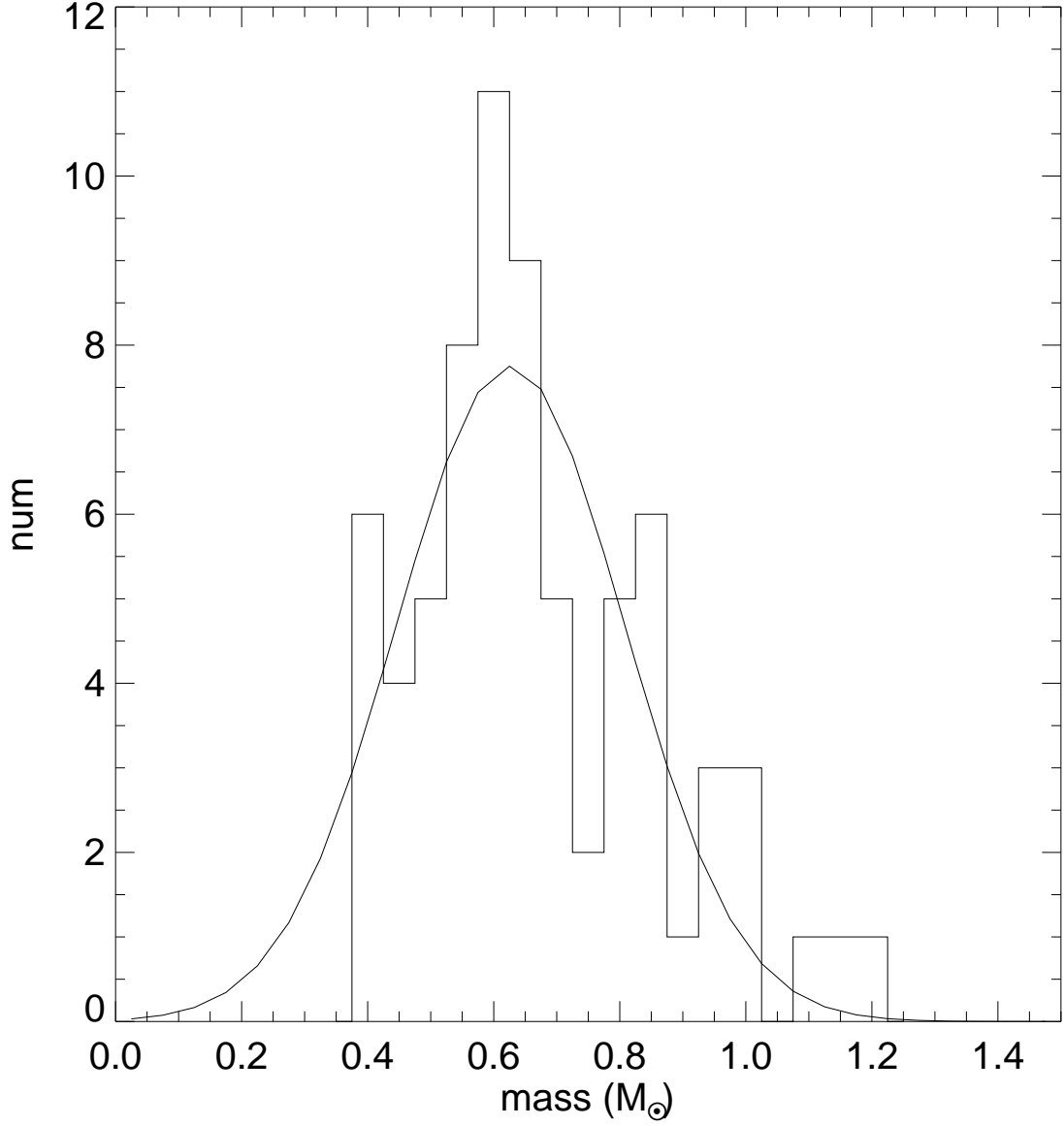}
\caption{Mass distribution of our sample of candidate WDs. The curve is a Gaussian fit with a peak at about 0.61 $M_{\odot}$}
\end{figure}

\begin{figure}
\epsscale{1.0}
\plotone{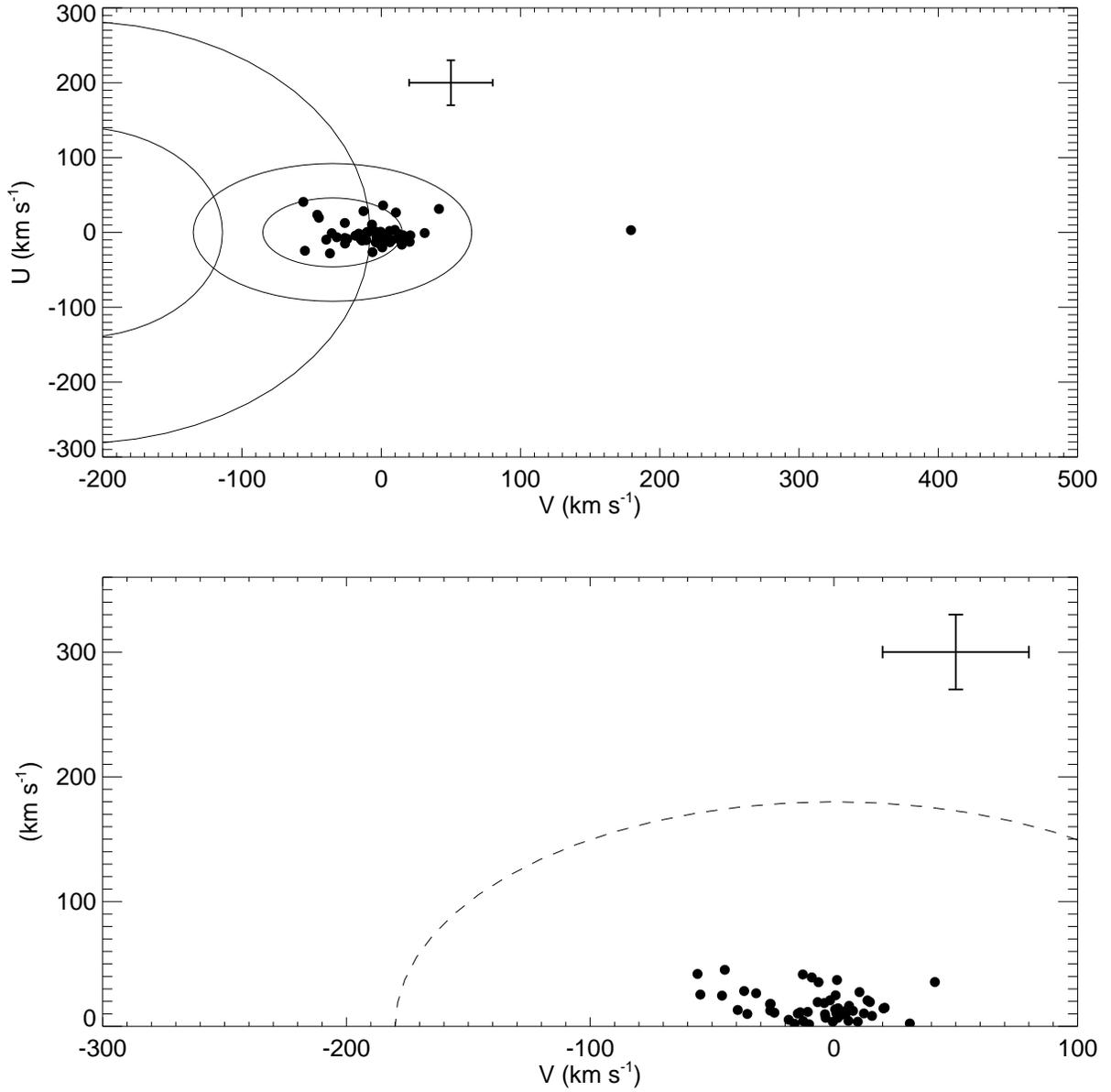}
\caption{Top: UV-velocity distribution of our WD candidates. Ellipsoids indicate 1$\sigma$ (inner) and 2$\sigma$ (outer) contours
for Galactic thick disk and halo populations, respectively. Bottom: Toomre diagram of our WDs. Dashed line
is V$\rm_{total}$ = 180 km s$^{-1}$.}
\end{figure}

\begin{figure}
\epsscale{1.0}
\plotone{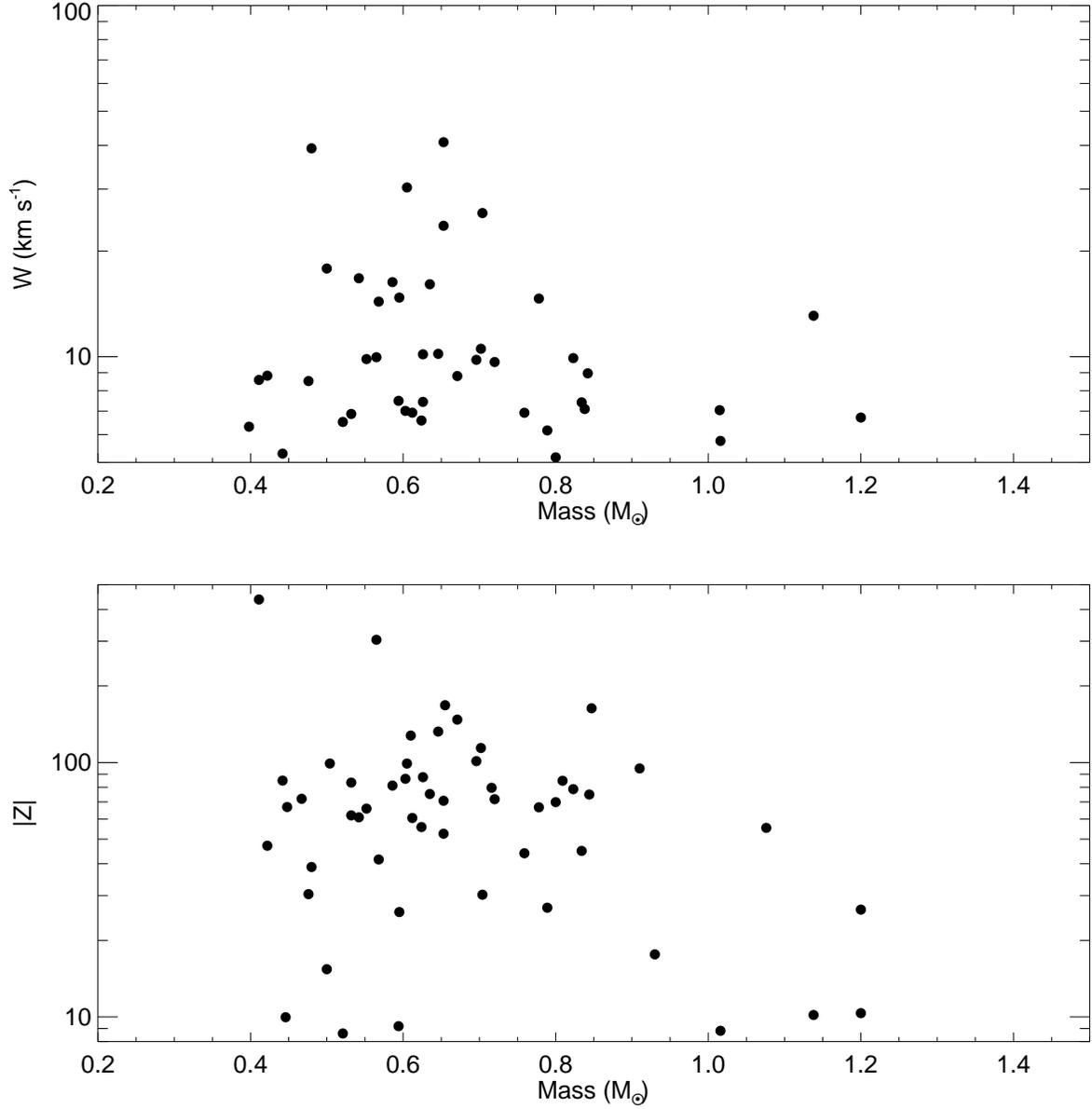}
\caption{Top: W vs. Mass. Bottom: $|$Z$|$ vs. Mass. Lower mass WDs clearly tend to have larger dispersion in both W velocity and vertical distance of the Galactic plane $|$Z$|$. }
\end{figure}

\clearpage








\clearpage


\begin{deluxetable}{ccccccccccccccccc}
\tablecolumns{16}
\tabletypesize{\scriptsize}
\rotate
\setlength{\tabcolsep}{0.02in}
\tablewidth{1000pt}
\tablecaption{catalog of DA white dwarfs.}
\tablehead{
\colhead{No}&\colhead{LAMOST Obj} &  \colhead{RA}   &
\colhead{DEC} & \colhead{T$\rm_{eff}$}      & \colhead{log $g$}     & \colhead{mass}    & \colhead{age}&    \colhead{u}&
 \colhead{g}   & \colhead{r}& \colhead{i}& \colhead{z}& \colhead{V} & \colhead{source}\tablenotemark{a} & \colhead{E(B-V)}& \colhead{dis}\\
 &&(deg)&(deg)&(K)&&(M$_{\odot}$&(Myr)&&&&&&&&(pc)}
\startdata
 0  &J220522.86+021837.56  & 331.345250  &   2.310432  &   15377  $\pm$     493      &     8.02  $\pm$     0.10      &     0.63  $\pm$     0.06      &     190  $\pm$      43      &    17.35  &    17.00  &    17.25  &    17.45  &    17.71  &       &1&     0.05  &   135   \\
 1&J025737.25+264047.89  &  44.405201  &  26.679970  &   19008  $\pm$     669      &     7.87  $\pm$     0.12      &     0.55  $\pm$     0.06      &      66  $\pm$      21      &      &    16.91  &    17.00  &    17.16  &      & &2     &     0.16  &   139   \\
 2&J030214.72+285707.41  &  45.561340  &  28.952057  &   21894  $\pm$    1406      &     8.01  $\pm$     0.23      &     0.64  $\pm$     0.13      &      46  $\pm$      37      &     &    17.21  &    17.60  &    17.80  &      &  &2    &     0.18  &   173   \\
 3&J040449.34+280023.65  &  61.205600  &  28.006570  &   29302  $\pm$    2525      &     8.25  $\pm$     0.55      &     0.79  $\pm$     0.33      &      20  $\pm$      32      &     &    15.96  &    15.84  &    16.00  &      &   &2   &     0.21  &    87   \\
 4&J004036.79+413138.79  &  10.153296  &  41.527443  &   13000  $\pm$     651      &     7.75  $\pm$     0.08      &     0.48  $\pm$     0.04      &     216  $\pm$      47      &      &    15.90  &    16.21  &    16.40  &      &   &2   &     0.07  &    83   \\
 5&J003956.55+422929.55  &   9.985629  &  42.491542  &   18053  $\pm$     816      &     7.32  $\pm$     0.15      &     0.34  $\pm$     0.05      &      50  $\pm$      10      &      &    16.43  &    16.58  &    16.72  &      &   &2   &     0.06  &   181   \\
 6&J004128.67+402324.09  &  10.369458  &  40.390026  &   25996  $\pm$     733.      &     7.92  $\pm$     0.10      &     0.59  $\pm$     0.05      &      15.  $\pm$       3.      &       &       &      &       &       &    17.14  & 3&    0.08  &    68   \\
 7&J005340.53+360116.89  &  13.418857  &  36.021358  &   29772  $\pm$     158      &     7.96  $\pm$     0.04      &     0.63  $\pm$     0.02      &       9  $\pm$       0      &     &    14.10  &    14.58  &    14.91  &      &    &2  &     0.05  &    72   \\
 8&J100551.51-023417.87  & 151.464628  &  -2.571630  &   22072  $\pm$     477      &     8.22  $\pm$     0.07      &     0.76  $\pm$     0.04      &      78  $\pm$      16      &    15.15  &    15.10  &    15.46  &    15.76  &    16.08  &   &1   &     0.05  &    68   \\
 9&J100316.35-002336.95  & 150.818141  &  -0.393597  &   22249  $\pm$     330      &     7.92  $\pm$     0.05      &     0.59  $\pm$     0.03      &      33  $\pm$       4      &    15.97  &    15.93  &    16.25  &    16.56  &    16.85  &  &1    &     0.05  &   123   \\
10  &J100941.45-004404.55  & 152.422705  &  -0.734597  &   16489  $\pm$     601      &     7.98  $\pm$     0.13      &     0.60  $\pm$     0.08      &     140  $\pm$      45      &    17.36  &    16.98  &    17.24  &    17.44  &    17.74  &&1      &     0.04  &   148   \\
11  &J054613.53+255031.70  &  86.556364  &  25.842139  &   22935  $\pm$     498      &     7.99  $\pm$     0.07      &     0.63  $\pm$     0.04      &      34  $\pm$      10      &      &    17.33  &    17.62  &    17.78  &      &      & 2 &   1.72  &    27   \\
12  &J090734.26+273903.32  & 136.892757  &  27.650923  &   18619  $\pm$     386      &     8.56  $\pm$     0.07      &     0.97  $\pm$     0.04      &     272  $\pm$      39      &    16.31  &    16.08  &    16.37  &    16.64  &    16.89  &      & 1&    0.03  &    72   \\
13  &J004628.31+343319.90  &  11.617971  &  34.555527  &   14644  $\pm$     808      &     7.60  $\pm$     0.18      &     0.41  $\pm$     0.08      &     120  $\pm$      47      &    16.83  &    16.33  &    16.40  &    16.53  &    16.75  &      & 1&    0.08  &   112   \\
14  &J005340.53+360116.89  &  13.418857  &  36.021358  &   26534  $\pm$     394      &     7.88  $\pm$     0.06      &     0.58  $\pm$     0.03      &      13  $\pm$       1      &      &    14.10  &    14.58  &    14.91  &      & &2     &     0.05  &    67   \\
15  &J052038.36+304822.65  &  80.159836  &  30.806293  &   15924  $\pm$     348      &     8.00  $\pm$     0.07      &     0.61  $\pm$     0.04      &     164  $\pm$      28      &      &    15.38  &    15.68  &    15.88  &      & &2     &     0.85  &    24   \\
16  &J031236.50+515511.74  &  48.152099  &  51.919927  &   23558  $\pm$    1966      &     7.93  $\pm$     0.29      &     0.59  $\pm$     0.14      &      25  $\pm$      13      &      &      &      &      &      &    15.44  & 3&    0.84  & 103  \\
17  &J055046.51+261220.27  &  87.693772  &  26.205631  &   28000  $\pm$    1916      &     8.34  $\pm$     0.39      &     0.84  $\pm$     0.24      &      37  $\pm$      57      &      &    15.13  &    15.64  &    15.91  &      &&2      &     1.50  &    13   \\
18  &J013938.94+291859.80  &  24.912266  &  29.316611  &   20934  $\pm$     515      &     8.13  $\pm$     0.08      &     0.70  $\pm$     0.05      &      77  $\pm$      22      &      &    17.53  &    17.94  &    18.19  &      & &2     &     0.05  &   213   \\
19&J105811.27+475752.75  & 164.546942  &  47.964653  &   29532  $\pm$     490      &     7.84  $\pm$     0.11      &     0.56  $\pm$     0.05    &       9  $\pm$       0      &    17.09  &    17.29  &    17.75  &    18.10  &    18.35  & &1     &     0.01  &   353   \\
20  &J104311.45+490224.35  & 160.797708  &  49.040097  &            &                &                 &           &    15.47  &    15.84  &    16.40  &    16.76  &    17.18  &      &1&     0.01  &       \\
21  &J053931.86+285456.66  &  84.882770  &  28.915740  &   23865  $\pm$    1774      &     8.63  $\pm$     0.26      &     1.01  $\pm$     0.15      &     147  $\pm$      97      &      &    17.39  &    16.64  &    16.17  &      & &2     &     1.43  &    15  \\
22  &J094104.43+282224.58  & 145.268457  &  28.373495  &   16713  $\pm$     438      &     7.86  $\pm$     0.09      &     0.54  $\pm$     0.05      &     109  $\pm$      24      &    15.70  &    15.42  &    15.70  &    15.94  &   16.25  &      &1&     0.02  &    82  \\
23  &J081845.28+121952.45  & 124.688667  &  12.331236  &   22271  $\pm$     531      &     8.34  $\pm$     0.08      &     0.83  $\pm$     0.05      &     100  $\pm$      22      &    16.32  &    16.18  &    16.57  &    16.88  &    17.19  &      & 1&    0.03  &   107   \\
24  &J014147.59+302135.45  &  25.448307  &  30.359846  &   17520  $\pm$     367      &     8.17  $\pm$     0.07      &     0.72  $\pm$     0.04      &     162  $\pm$      27      &      &    16.96  &    17.39  &    17.54  &      &  &2    &     0.05  &   138   \\
25  &J014933.76+285610.60  &  27.390679  &  28.936279  &   32200  $\pm$     631      &     8.33  $\pm$     0.12      &     0.84  $\pm$     0.07      &      17  $\pm$      10      &      &    16.88  &    17.47  &       &      &      &2&     0.06  &   140   \\
26  &J074742.05+280945.57  & 116.925192  &  28.162658  &   15085  $\pm$     596      &     7.66  $\pm$     0.13      &     0.44  $\pm$     0.06      &     117  $\pm$      31      &    17.83  &    17.43  &    17.69  &    17.88  &    18.13  &    &1  &     0.04  &   209   \\
27  &J075251.35+271513.85  & 118.213962  &  27.253847  &   25134  $\pm$     711      &     7.94  $\pm$     0.10      &     0.60  $\pm$     0.06      &      19  $\pm$       9      &    16.72  &    16.73  &    17.15  &    17.46  &    17.79  &      & 1&    0.03  &   206   \\
28  &J075106.48+301726.96  & 117.776979  &  30.290822  &   34418  $\pm$     580      &     8.21  $\pm$     0.10      &     0.78  $\pm$     0.06      &       9  $\pm$       1      &    15.65  &    15.92  &    16.39  &    16.72  &    17.05  &      &1&     0.05  &   156   \\
29&J113614.04+290130.26  & 174.058504  &  29.025072  &   24106  $\pm$     255      &     7.80  $\pm$     0.03      &     0.53  $\pm$     0.02    &      20  $\pm$       1      &    14.64  &    14.68  &    15.13  &    15.44  &    15.75  &      &1&     0.02  &    87   \\
30  &J113705.17+294757.54  & 174.271529  &  29.799317  &   21786  $\pm$     160      &     8.58  $\pm$     0.03      &     0.98  $\pm$     0.02     &     174  $\pm$      11      &    12.29  &    12.31  &    12.69  &    12.99  &    13.31  &      &1&     0.02  &    15   \\
31  &J113423.35+314606.58  & 173.597300  &  31.768494  &   14683  $\pm$     832      &     8.02  $\pm$     0.14      &     0.62  $\pm$     0.08      &     219  $\pm$      73      &    15.53  &    15.17  &    15.44  &    15.68  &    15.95  &    &1  &     0.03  &    58  \\
32  &J093903.33+114418.62  & 144.763879  &  11.738506  &   16673  $\pm$     815      &     8.75  $\pm$     0.09      &     1.08  $\pm$     0.05      &     513  $\pm$     116      &    17.37  &    17.01  &    17.21  &    17.41  &    17.67  &   &1   &     0.03  &    82   \\
33  &J070755.01+265102.94  & 106.979210  &  26.850817  &   17854  $\pm$     893      &     8.87  $\pm$     0.12      &     1.14  $\pm$     0.06      &     554  $\pm$     202      &      &    15.53  &    15.86  &    16.01  &      & &2     &     0.07  &    39   \\
34  &J104946.47+003635.81  & 162.443625  &   0.609947  &   19832  $\pm$     550      &     8.08  $\pm$     0.10      &     0.67  $\pm$     0.06      &      87  $\pm$      23      &    17.25  &    17.27  &    17.67  &    17.97  &    18.30  &      &1&     0.05  &   191   \\
35  &J104623.28+024236.57  & 161.596987  &   2.710158  &   13000  $\pm$     728      &     7.73  $\pm$     0.08      &     0.47  $\pm$     0.04      &     211  $\pm$      52      &    16.43  &    16.03  &    16.26  &    16.48  &    16.72  &      &1&     0.04  &    92   \\
36  &J104928.89+275423.77  & 162.370375  &  27.906603  &   14212  $\pm$     681      &     7.68  $\pm$     0.15      &     0.45  $\pm$     0.07      &     148  $\pm$      44      &    15.74  &    15.32  &    15.51  &    15.75  &    15.98  &      &1&     0.02  &    75   \\
37  &J115506.22+264924.59  & 178.775929  &  26.823497  &   17291  $\pm$     679      &     8.47  $\pm$     0.13      &     0.91  $\pm$     0.08      &     285  $\pm$      88      &    17.03  &    16.69  &    16.97  &    17.23  &    17.50  &      &1&     0.02  &    97   \\
38  &J094627.81+313211.08  & 146.615867  &  31.536411  &   15000  $\pm$    2362      &     8.34  $\pm$     0.20      &     0.82  $\pm$     0.13      &     342  $\pm$     218      &    17.30  &    16.91  &    17.12  &    17.34  &    17.55  &     &1&     0.02  &   103   \\
39&J070057.53+284310.06  & 105.239692  &  28.719461  &   16000  $\pm$     735      &     8.15  $\pm$     0.13      &     0.70  $\pm$     0.08    &     207  $\pm$      66      &    17.37  &    16.98  &    17.23  &    17.45  &    17.73  &      &1&     0.08  &   120   \\
40  &J040613.25+465133.66  &  61.555205  &  46.859349  &   33026  $\pm$     436      &     7.50  $\pm$     0.10      &     0.45  $\pm$     0.03      &       6  $\pm$       1      &      &      &      &      &      &    14.77  & 3&    0.82  & 145  \\
41  &J103535.22+395502.27  & 158.896764  &  39.917298  &   16652  $\pm$     550      &     8.05  $\pm$     0.11      &     0.65  $\pm$     0.07      &     155  $\pm$      39      &    17.43  &    17.12  &    17.33  &    17.55  &    17.84  &     &1&     0.01  &   154   \\
42  &J105443.36+270658.42  & 163.680650  &  27.116228  &   24915  $\pm$     131      &     8.38  $\pm$     0.02      &     0.86  $\pm$     0.02      &      74  $\pm$       4      &    13.86  &    13.98  &    14.34  &    14.64  &    14.97  &      &1&     0.02  &    41   \\
43  &J064452.84+260947.75  & 101.220170  &  26.163263  &   16835  $\pm$     598      &     7.78  $\pm$     0.13      &     0.50  $\pm$     0.06      &      90  $\pm$      20      &      &    15.48  &    15.98  &    16.22  &      &  &2    &     0.10  &    86   \\
44  &J065601.55+115745.85  & 104.006460  &  11.962736  &   31347  $\pm$     603      &     7.42  $\pm$     0.15      &     0.41  $\pm$     0.05      &       8  $\pm$       1      &      &    14.31  &    13.48  &    13.13  &    11.94  &      & 1&    0.22  &    63   \\
45  &J013914.45+290057.61  &  24.810197  &  29.016003  &   16808  $\pm$     478      &     8.06  $\pm$     0.10      &     0.65  $\pm$     0.06      &     153  $\pm$      34      &      &    16.20  &    16.53  &    16.68  &      & &2     &     0.06  &    97   \\
46  &J094126.79+294503.39  & 145.361630  &  29.750942  &   21798  $\pm$     267      &     8.15  $\pm$     0.04      &     0.72  $\pm$     0.03      &      68  $\pm$      10      &    15.88  &    15.91  &    16.25  &    16.55  &    16.87  &      &  1&   0.02  &   106   \\
47  &J100549.01+424804.68  & 151.454200  &  42.801300  &   23923  $\pm$     812      &     8.11  $\pm$     0.12      &     0.70  $\pm$     0.07      &      38  $\pm$      13      &    16.04  &    16.00  &    16.39  &    16.70  &    16.98  &      & 1&    0.01  &   127   \\
48  &J093047.11+160012.98  & 142.696300  &  16.003606  &   32492  $\pm$     634      &     8.00  $\pm$     0.13      &     0.66  $\pm$     0.07      &       7  $\pm$       1      &    16.35  &    16.64  &    17.12  &    17.50  &    17.84  &      & 1&    0.04  &   249   \\
49&J093451.69+171814.00  & 143.715358  &  17.303889  &   14645  $\pm$     566      &     7.80  $\pm$     0.12      &     0.50  $\pm$     0.06    &     156  $\pm$      45      &    17.10  &    16.79  &    17.09  &    17.36  &    17.66  &      & 1&    0.03  &   143   \\
50  &J092518.36+180534.20  & 141.326500  &  18.092833  &   26274  $\pm$     324      &     8.29  $\pm$     0.06      &     0.81  $\pm$     0.04      &      43  $\pm$      11      &    16.07  &    16.17  &    16.61  &    16.94  &    17.23  &      &1&     0.05  &   127   \\
51  &J052147.24+283532.50  &  80.446823  &  28.592361  &   18917  $\pm$     466      &     7.81  $\pm$     0.09      &     0.52  $\pm$     0.05      &      59  $\pm$      16      &      &    17.50  &    17.73  &    17.86  &      & &2     &     0.62  &   108   \\
52  &J071223.81+260933.41  & 108.099190  &  26.159281  &   14278  $\pm$     632      &     7.75  $\pm$     0.14      &     0.48  $\pm$     0.06      &     159  $\pm$      40      &      &    16.85  &    17.24  &    17.41  &      &  &2    &     0.08  &   141   \\
53  &J102521.36+455553.91  & 156.338987  &  45.931643  &   23547  $\pm$     908      &     7.51  $\pm$     0.13      &     0.41  $\pm$     0.05      &      19  $\pm$       3      &    18.12  &    18.27  &    18.60  &    18.89  &    19.26  &      & 1&    0.02  &   530   \\
54  &J101806.60+455830.36  & 154.527482  &  45.975101  &   22475  $\pm$    1541      &     8.36  $\pm$     0.22      &     0.85  $\pm$     0.14      &     101  $\pm$      59      &    17.61  &    17.55  &    17.90  &    18.19  &    18.47  &      & 1&    0.01  &   201   \\
55  &J033149.69+305944.92  &  52.957023  &  30.995811  &   19435  $\pm$     332      &     8.64  $\pm$     0.06      &     1.02  $\pm$     0.03      &    277  $\pm$      37      &      &    16.89  &    17.41  &    17.67  &      &   &2   &     1.18  &    25   \\
56  &J033253.91+284006.91  &  53.224625  &  28.668586  &   19000  $\pm$    1400      &     9.84  $\pm$     0.09      &     1.20  $\pm$     0.15      &     155  $\pm$       0      &      &    17.06  &    17.41  &    17.58  &      &   &2   &     0.25  &    27   \\
57  &J090918.99+292929.61  & 137.329125  &  29.491558  &   22588  $\pm$     346      &     8.04  $\pm$     0.05      &     0.65  $\pm$     0.03      &      43  $\pm$       8      &    15.74  &    15.68  &    16.04  &    16.34  &    16.59  &      &1&     0.02  &   107   \\
58  &J102155.50+405014.85  & 155.481261  &  40.837458  &   23364  $\pm$     999      &     7.96  $\pm$     0.15      &     0.61  $\pm$     0.09      &      28  $\pm$      18      &    16.19  &    16.25  &    16.61  &    16.91  &    17.23  &      &1&     0.01  &   153   \\
59&J121336.54+314808.77  & 183.402250  &  31.802436  &   13308  $\pm$     405      &     8.26  $\pm$     0.08      &     0.77  $\pm$     0.05    &     418  $\pm$      67      &    16.21  &    15.79  &    16.00  &    16.21  &    16.49  &      &1&     0.01  &    60   \\
60  &J134922.51-003503.15  & 207.343783  &  -0.584208  &   16401  $\pm$    1151      &     8.52  $\pm$     0.19      &     0.94  $\pm$     0.12      &     363  $\pm$     153      &    17.24  &    16.91  &    17.25  &    17.46  &    17.73  &      &1&     0.03  &    99  \\
61  &J144433.83-005958.83  & 221.140967  &  -0.999675  &   12165  $\pm$     856      &     8.01  $\pm$     0.21      &     0.61  $\pm$     0.13      &     367  $\pm$     154      &    16.58  &    16.20  &    16.38  &    16.58  &    16.85  &      & 1&    0.04  &    77   \\
62  &J112518.85+541936.65  & 171.328550  &  54.326847  &   15272  $\pm$     209      &     7.85  $\pm$     0.04      &     0.53  $\pm$     0.02      &     147  $\pm$      14      &    15.62  &    15.28  &    15.57  &    15.83  &    16.08  &      & 1&    0.01  &    73   \\
63  &J113203.47+065509.52  & 173.014441  &   6.919311  &   12455  $\pm$    1141      &     7.93  $\pm$     0.45      &     0.57  $\pm$     0.25      &     310  $\pm$     201      &    15.21  &    14.89  &    15.13  &    15.35  &    15.64  &      &1&     0.04  &    46   \\
64  &J084107.69+163221.71  & 130.282053  &  16.539363  &   16626  $\pm$     580      &     8.30  $\pm$     0.11      &     0.80  $\pm$     0.07      &     237  $\pm$      64      &    17.52  &    17.22  &    17.45  &    17.67  &    17.97  &      & 1&    0.02  &   134   \\
65  &J150156.26+302300.13  & 225.484400  &  30.383369  &   27051  $\pm$     339      &     7.84 $\pm$     0.05      &     0.56  $\pm$     0.03      &      12  $\pm$       1      &    14.13  &    14.24  &    14.71  &    14.96  &    15.32  &      & 1&    0.02  &    77   \\
66  &J063406.26+025401.30  &  98.526074  &   2.900361  &   31607  $\pm$    1274     &     7.82  $\pm$     0.31      &     0.56  $\pm$     0.14      &       7  $\pm$       1      &      &      &      &      &      &     &4 &     1.60  &   \\
67  &J064438.16+030704.39  & 101.159020  &   3.117885  &   12067  $\pm$    2288      &     8.37  $\pm$     0.65      &     0.84  $\pm$     0.42      &     650  $\pm$     791      &      &    14.10  &    13.66  &    13.51  &      & &2    &     1.02  &     5   \\
68  &J063517.47+054917.94  &  98.822796  &   5.821650  &   22885.  $\pm$    3739      &     7.48  $\pm$     0.54      &     0.40  $\pm$     0.14      &      21  $\pm$      13      &     &    14.19  &    13.96  &    13.84  &      &  &2    &     0.41  &    38   \\
69&J152130.83-003055.70  & 230.378443  &  -0.515472  &   13000  $\pm$    1056      &     7.63  $\pm$     0.10      &     0.42  $\pm$     0.05    &     186.  $\pm$      71.      &    15.52  &    15.24  &    15.53  &    15.78  &    16.09  &      &  1&   0.07  &    67   \\
70  &J113705.14+294757.77  & 174.271408  &  29.799381  &   23829  $\pm$     127      &     8.49  $\pm$     0.02      &     0.93  $\pm$     0.02      &     111  $\pm$       6      &    13.54  &    12.45  &    12.88  &    13.16  &    13.69  &      & 1&    0.02  &    18   \\
71  &J191927.67+395839.30  & 289.865292  &  39.977583  &   20376  $\pm$     345      &     7.93  $\pm$     0.06      &     0.59  $\pm$     0.03      &      54  $\pm$       8      &      &      &      &      &      &      &4&     0.14  &  \\
\enddata
\tablenotetext{a}{1: SDSS 2: Xuyi Schmidt Telescope Photometric Survey of the Galactic Anti-center 3: UCAC 4: Kepler }


\end{deluxetable}




\end{document}